# A first approach for a possible cellular automaton model of fluids dynamics

Gianluca Argentini
gianluca.argentini@riellogroup.com

*New Technologies & Models*
*Information & Communication Technology Department*
*Riello Group, Legnago (Verona), Italy*

February 2003

**Abstract**

*In this paper I present a first attempt for a possible description of fluids dynamics by mean of a cellular automata technique. With the use of simple and elementary rules, based on random behaviour either, the model permits to obtain the evolution in time for a two-dimensional grid, where one molecule of the material fluid can ideally place itself on a single geometric square. By mean of computational simulations, some realistic effects, here showed by use of digital pictures, have been obtained. In a subsequent step of this work I think to use a parallel program for a high performances computational simulation, for increasing the degree of realism of the digital rendering by mean of a three-dimensional grid too. For the execution of the simulations, numerical methods of resolution for differential equations have not been used.*

## 1. Introduction

In a rectangular grid of *n* x *m* identical squares we can consider, at one of the sides, from the initial time $t_0 = 0$ a constant flow of coloured geometric particles, representing the molecules of a real material fluid. The particles slide without friction on the grid.

At every temporal step $t_1 = 1$, $t_2 = 2$, ...., $t_N = N$, a molecule goes forward on the next row on the basis of some predefined rules:

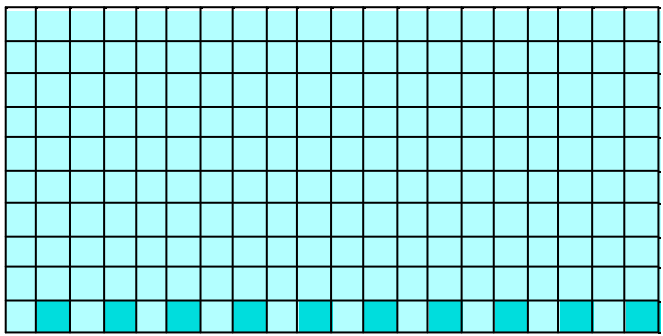

1. the motion can be toward the upper left, central or right cell, by mean of a random choice;
2. if the destination cell is already occupied by a predefined max number of molecules,

the particle remains on its own row, with the same previous random choice as motion direction;

3. if the destination cell is on the contrary occupied by an artificial obstacle, the particle remains on its own row, with the same previous random choice as motion direction too.

These rules now described are one of the possible evolution modes for a Cellular Automaton (CA), a logical instrument invented by John Von Neumann ([3]), widely used by Stephen Wolfram in his researches on complex dynamical systems ([4]) and by many experts of the international scientific community. The use of cellular automata for the representation of the evolution of a phenomenon, natural or not, presents two benefits respect a possible analytical description:

- *the facility of a computational implementation;*
- *the possibility of avoiding the formulation of a sophisticated mathematical model and/or the resolution of complicated differential equations, if they have been already found.*

Three examples explaining the results that one can obtain by mean of cellular automata techniques in the description of complex phenomena are offered by a simple model for the fluidodynamic behaviour ([4]), by a study on the liquids percolation ([1]) and by an elementary but effective scheme for the temporal evolution of seismic phenomena ([2]).

In this work I describe a first, very simple and not complete, model for simulating the dynamical behaviour of a fluid in presence of obstacles and I present a computational implementation. The three rules above mentioned have been used for obtain a first graphic representation. In a subsequent paper I'll present some qualitative and quantitative extensions of the model. We are developing some tests with the parallel systems of CINECA, Bologna (Italy) ([5]), for a high performances massive computational simulation.

## 2. The rules of the cellular automaton

We denote with $u_{i,j}(t_k)$ the number associated to the position (i, j), *j-th* place of *i-th* row, of the computational grid at time $t_k$. If $u_{i,j}(t_k) = 0$, no molecules are present in the position; if $u_{i,j}(t_k) = p > 0$ the position should be considered as occupied by p molecules, which we think as physically placed one upon the other on the plain of the grid. We assume that the time variable is incremented by discrete unitary values, so that each row of fluid molecules goes forward to the next at the same instant, for every occupied position and for every row of the grid. The rules which form the model of the used cellular automaton are schematically so described:

1. If $u_{i,j}(t_k) = p$, every molecule in the position (i, j) can move in the new position (i + 1, j + r) at the time instant $t_{k+1}$, where *r* is an integer randomly selected between the values –1, 0 and 1;

2. If $u_{i+1,j+r}(t_{k+1}) > d$, where *d* is the max allowed number of molecules in one single cell (darker tonality of blue in the picture) for advancing to the next row, the molecule will stay in the *i-th* row and it will move into the position (i, j + r);

3. If $u_{i+1,j+r}(t_{k+1}) < 0$, in the cell (i + 1, j + r) there is an obstacle, therefore no molecule can move into that position; the molecule will stay in the *i-th* row and it will move into the position (i, j + r).

The rule 1 imposes a constraint to the direction and velocity of molecules; the rule 2 doesn't allow the presence of material accumulation points for the molecules, and the rule 3 is a simple constraint of physical kind. In this first model we don't consider interaction forces of particular nature between the molecules or external forces acting on the fluid. Furthermore the molecules are considered as massless. From one of the sides of the grid we suppose the presence of a time-constant flow of new molecules, whose motion is always regulated by the same laws. The displacement of the molecules of *(i-1)-th* row is computed after that of the molecules of the *i-th* row, but ideally the rules are applied to every row of the grid at the same instant of time.

## 3. The computational algorithm

In the computational simulation, at start the global variables are defined and initialized. The positions grid is represented by a two-dimensional array *n* x *m*, while a variable of integer type is used to keep the maximum number of molecules which prevents the motion of a particle forward the next row; this number, in order to obtain a simulated approximation of the phenomenon of turbulence, doesn't set a superior limitation for the displacement on the same row where a molecule is already placed. A monodimensional array, appropriately initialized according to the experiment to be performed, is used for representing the initial positions of the molecules in the flow coming into the grid. Another monodimensional *carry* array is needed for registering the molecules of *k-th* row which, during a temporal evolution step, have a motion along the horizontal direction, because the positions in this manner occupied affect the evolution of the material *(k-1)-th* row at the same computation step.

The evolution of the two-dimensional grid during a single step begins from the first row of the free material front, which is placed p.e. at the *p-th* grid row, and finishes at the first row, that of the incoming flow. The two-dimensional array of the positions is on each step saved into an array which stores all the history of configurations of the previous steps, in order to obtain useful informations on the temporal dynamics and a possible graphic animation.

The main core of the algorithm for the computing of the new configuration of a material row (MR) begins with the assignment to a temporary array (TA) of the values of the carry array (YA) obtained by the computation on the previous row. The temporary array is used for assign the positions of the molecules during the running step. The array YA is then resetted. For every cell (i, j) occupied by a material particle of MR, one computes the new position of every molecule placed on that cell. Fixed a molecule, by mean of a random procedure one selects a number *r* from the set {-1, 0, 1}, with the necessary variations if the considered molecule is placed on a cell of the right or left side of the grid. If in the new position (i + 1, j + r) the number of molecules is equal or greater then the maximum limitation permitted for the advance to the new row, the molecule is moved to the position (i, j + r) of the array YA, therefore it remains in the same row where is already placed. If on the contrary the new position (i + 1, j + r) is physically occupied by an obstacle, the molecule is moved to the position (i, j + r) of the array YA too. In the remaining possibility, the molecule is moved to the position (i +1, j + r) of the array TA, and the value `u`$_{i+1,j+r}$ is incremented by one unit. When all the cell of MR have been so analyzed, the algorithm returns the value of TA to the calling procedure of the main program. Then by mean of graphic processing one obtain some pictures of the final grid, with the cells occupied by molecules coloured by blue tone of intensity proportional to the number of particles present on the cell.

## 4. First results

I present now some results obtained by mean of these first simple experiments.

The Fig.1 shows the final configuration of a 20 x 20 grid after 200 computational steps, when the incoming flow is represented by a single molecule in the right lower cell. A material obstacle is drawn by black color. For comfort of view the geometric mesh of the single component squares has been removed. It's visible, by mean of a blue color increasing tone, the effect of the molecules thickening in proximity to the obstacle and the turning of black cells by some particles.

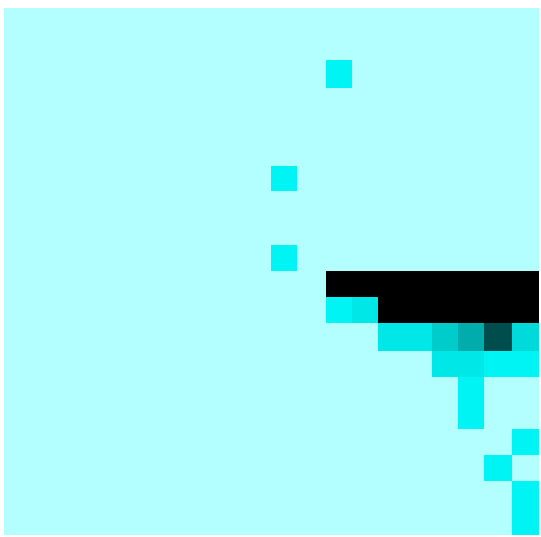

**Fig. 1**

The discrete degree of accordance with real situations can permit to deduce that the simple CA considered model could already be a first approximation for a more detailed and deeper description. The Fig. 2 shows the final configuration of a 100 x 100 grid after 100 computational steps, when the constant flow is represented by molecules incoming from cells geometrically alternated at the lower side. In this case too is well visible the effect of thickening of the molecules in proximity to the obstacle.

The Fig. 3 on the contrary shows the final configuration of a 100 x 100 grid after 200 computational steps, when the incoming flow is made by molecules at every cells of the lower side.

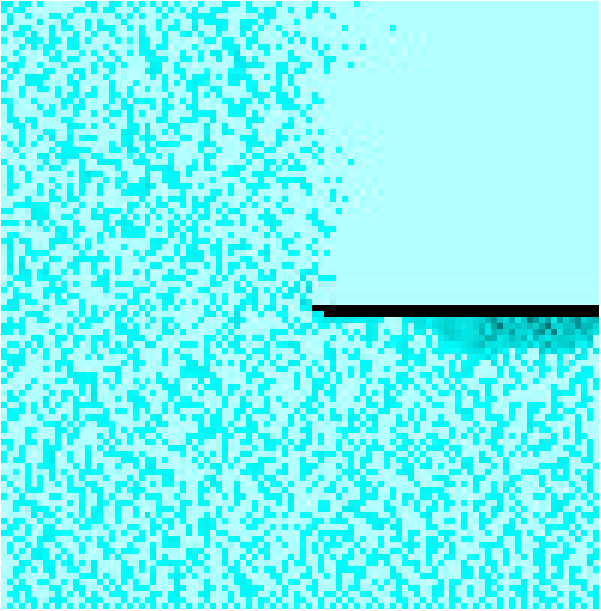
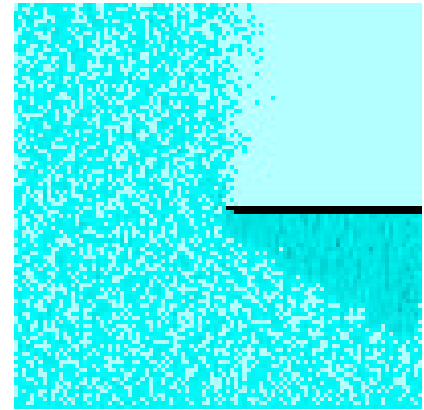

**Fig. 2** and **Fig. 3**

Since the number of steps respect to the preceding configuration is double, and having used the same number of molecules for cell as upper limitation for the advance to the next row, the material thickening due to the presence of obstacle shows a larger geometric extent but an equal local particles density respect to the preceding situation.

The Fig. 4 on the contrary is a three-dimensional representation of a 50 x 50 configuration, where the vertical height is determined by the number of molecules present on every cell of the base grid. The green wall is the obstacle, which thickens the particles in an evident fashion .

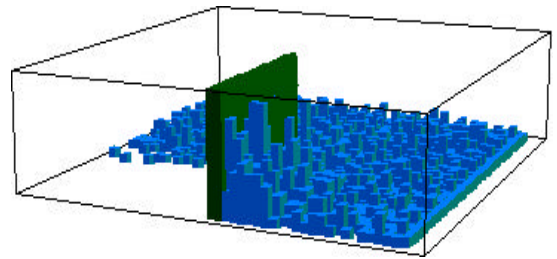

**Fig. 4**

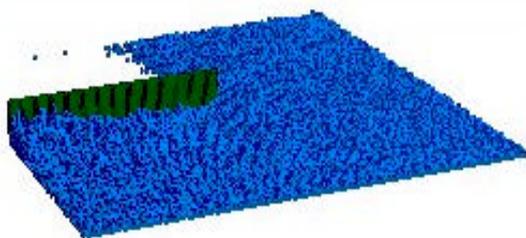

**Fig. 5**

The Fig. 5 is a three-dimensional representation of a 200 x 200 grid after 200 computational steps; besides a better resolution respect to the preceding picture, in the middle is visible the flow lines bending induced by the presence of the obstacle.

## 5. First considerations

The CA model upon described is very elementary and limited, but from the preceding pictures we can deduce that it provides a first useful approximation for the computational treatment of fluidodynamic phenomena. In particular the discrete level of realism obtained with the graphics of

Fig. 5 is not easily achievable by an analytic model as Navier-Stokes differential system; moreover the study of such mathematical model is not straightforward when one consider a flow in presence of bodies, like a generic obstacle, which are not component of the fluid ([4]).

We have made some quantitative observations on the parameter *d*, used in the model, maximum number of molecules per cell that limits the advance to the next row for a given particle. This value seems to give a strong contribution to determine the empiric value *max_mol*, maximum observed number of molecules which have thickened in some cell for the effect of the carries. In Fig.6 is drawn a graphic obtained with some tests on a 50 x 50 grid; one can notice a nearly logarithmic dependence (red curve) of *max_mol* (black line) respect to the parameter *d*.

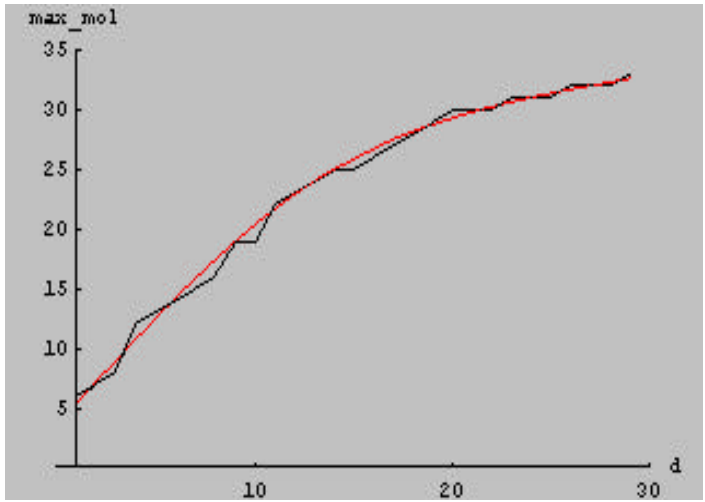

Fig. 6

A similar logarithmic behaviour seems to bind the value *max_mol* to the horizontal dimension of the grid and to the linear length of the used obstacle. Since *d* can be assimilated to a density parameter, one can deduce that *max_mol* seems to have the same characteristics of the *Reynolds number* ([4]) relating to the considered physical situation. The molecules thickening in a cell can be interpreted as an index of turbulence in that position, therefore an high value for *max_mol* seems to express the same physical meaning of the Reynolds number.

The possibility of a mathematical relation between the two quantities is under inquiry too.

### 6. Possible extensions of the model

The used CA model can be improved for obtain a greater quantity of informations, a better qualitative description of the investigated dynamics and a more realistic graphics. Some possible extensions at present under study are:

- the possibility for the molecules of movement into the previous row respect to that where they are placed and the possible advance toward a row beyond that one geometrically next;
- the generalization of the model to 3D;
- the use of a high performance computational simulation by mean of parallel libraries, as MPI ([6]), and parallel computer, as the cluster Linux of Cineca at Bologna (Italy), for the purpose of testing high values for the geometric dimension of the grid and for the molecules number of the flow, so obtaining realistic graphic pictures;
- the insertion of new rules for treating material mass and intermolecular forces effects;

- the use of more general geometric shapes both for the cells grid and for the obstacles inner the fluid;
- the research of a relation between the type of the used CA model and the phenomenon of fluidodynamic turbulence.

## 7. Technical notes

The used CA model has been simulated on a personal computer by mean of a program written in the language *Mathematica* of Wolfram Research. The two-dimensional graphics has been obtained by mean of the function `DensityGraphics` using a personalized version of the colouring procedure `RGBColor`. The three-dimensional graphics instead has been obtained by the use of the primitive `Cuboid`, which has permitted the drawing of cubes with the face equivalent to the grid cell and stacked for a number equal to that of the molecules present on the base cell; then the rendering has been obtained by the procedure `Graphics3D`, with the use of the function `ViewPoint` for determining the spatial view point.

For the implementation on Linux cluster we are using the *C* language interfaced with the parallelization and message passing library MPI.

## 8. Bibliography

[1] R. Cappuccio, G. Cattaneo, G. Erbacci, U. Jocher, *A parallel implementation of a cellular automata based model for coffee percolation*, Parallel Computing, vol. 27, n. 5, 2001.

[2] Silvia Castellaro, Francesco Mulargia, *A simple but effective cellular automaton for earthquakes*, Geophysical Journal International, 144, 2001.

[3] John Von Neumann, Arthur W. Burks, *Theory of Self-Reproducing Automata*, Univ. of Illinois Press, Urbana, 1966.

[4] Stephen Wolfram, *Fluid flow Notes*, in *A new kind of Science*, Wolfram Media, Inc., 2002.

[5] www.cineca.it, Web site of *Consorzio Interuniversitario Nord Est per il Calcolo Automatico*.

[6] www.mcs.anl.gov/mpich, Web site for *MPI* libraries.

The author wishes to thank dr. Paolo Malfetti and the staff Linux directed by dr. Stefano Martinelli for the possibility of using the 128-processors Beowulf Linux cluster of Cineca for the simulations tests. A thank also to dr. Stefano Cozzini of Democritos Centre of INFM, Istituto Nazionale di Fisica della Materia, Italy, and to dr. Carlo Cavazzoni for their useful lectures on High Performances Computing Linux clusters.

Gianluca Argentini
gianluca.argentini@riellogroup.com
febbraio 2003